\documentclass[aps,twocolumn,floatfix]{revtex4}
\usepackage{graphicx}
\usepackage{bm}

\newcommand{\Tr}{{\rm Tr}}
\newcommand{\fec}{\mathcal{F}_e\mathcal{(C)}}

\begin{document}

\title{The Falicov--Kimball model in external magnetic field: orbital effects} 
\author{  Maciej Wr\'obel,
  Marcin Mierzejewski, and
  Maciej M. Ma\'ska}
\email{maciek@phys.us.edu.pl}
\affiliation{%
Department of Theoretical Physics, Institute of Physics, University of Silesia, Katowice, Poland}

\begin{abstract}{ We study thermodynamic properties of the two--dimensional (2D) 
Falicov--Kimball model in the presence of external magnetic field perpendicular to the lattice. The field
 is taken into account by the Peierls
  substitution in the hopping term.
  In the non--interacting case the field
  dependent energy spectrum forms the famous Hofstadter butterfly. Our results indicate 
  that for arbitrary nonzero interaction strength and arbitrary magnetic field
  there is a gap in the energy spectrum at sufficiently
low temperature. The gap vanishes
  with increase of temperature for weak coupling, however, it persists
  at high temperatures if the coupling is strong enough. Numerical results have 
been obtained with the help of Monte Carlo technique based on a modified 
Metropolis algorithm.}
\end{abstract}
\maketitle

\section{Introduction}
Strongly correlated electron systems have generated interest over the
last few decades. It is widely accepted that many phenomena in
condensed matter physics are connected with electronic
correlations. Moreover an experimental progress made both in confining
quantum gases and in preparation of nanosystems lead to requirement of
theoretical models, that allow description of correlated quantum
systems in periodic potentials and under an influence of the external
magnetic field. Unfortunately, taking into account of all these three
factors simultaneously is intractable. While the problem of electrons
in a periodic potential under influence of the external magnetic field
has been investigated since the beginning of the quantum mechanic, its
solutions are known only in a few cases. In particular, one may prove
that in two dimensional fermionic gases orbital effects due to the
magnetic field lead the energy spectrum to form the famous Hofstadter
butterfly. Incorporation of the electron correlations in such a model
encounters yet unresolved problems, mostly because of lack of relevant
mathematical methods.

The Hubbard model is frequently used as a starting point for studying
strongly correlated systems. Originally used for a description the
metal--insulator transition, it reveals interesting solutions,
describing various phenomena observed in strongly correlated
systems. Despite the simplicity of the model, only a few rigorous
results are known, mostly in one- or infinite--dimensional cases. Other
results have been obtained with the help of approximate
methods. Some attempts have been made to analyze the orbital effects in
the Hubbard model in an external magnetic field. Some of them, based
on an exact diagonalization,\cite{CiG} were obtained for relatively
small systems and suffer from the finite size effects. The other
attempts based on the mean--field approximation (MFA)\cite{salk} are
questionable due to the limited applicability of the MFA in low dimensional 
systems.

The lack of exact solutions for the Hubbard model and ambiguity of
solutions obtained within approximate methods encouraged us to study even
simpler model, i.e., the Falicov--Kimball model. It was
proposed by Hubbard and Gutzwiller (for review, see, e.g.,
\cite{gruber-2005}, \cite{gruber-1998}) as a simplification of the
Hubbard model and further was redeveloped by Falicov and Kimball to
study phase transitions in rare earths and transition metals.
\cite{PhysRevLett.22.997}  The model is a limiting case of the
asymmetric one-band Hubbard model where the mass of spin--down electrons 
goes to infinity. It describes a system consisting
two kinds of fermions. One of them are itinerant particles and the
others are massive and therefore localized. The only
interaction in the Falicov--Kimball model is the on--site Coulomb repulsion 
between itinerant and localized particles. In the second quantization the 
model is described by the following Hamiltonian:
\begin{equation}
  \label{eq:ham}
  \mathcal{H}=-t\sum_{\langle i,j \rangle} c^\dagger_i c_j + U\sum_j f^\dagger_if_i c^\dagger_i c_i, 
\end{equation}
where $t$ is the hopping integral, $c^\dagger_i$ ($f^\dagger_i$) are the 
creation operators of an itinerant (localized) fermion at site $i$ and $U$ is the 
Coulomb interaction.

While the Falicov--Kimball model is much simpler than its predecessor,
it still cannot be rigorously solved in a general case. Fortunately,
a significantly larger number of exact results is known for the the
Falicov--Kimball model than for the Hubbard model.\cite{freericks-2003-75} 
One of the most important theorem, proved by Kennedy and Lieb,
\cite{KennedyAndLieb} states that at low enough temperature there is a long
range order for lattices of dimensionality greater than one for certain
fillings and for all values of $U$. It has been shown that the
ordering comes from effective correlations between states of one kind,
even though the model does not contain a direct interaction of this type. There are also important approximate results for the
Falicov--Kimball model (see, e.g., Ref. \cite{freericks-2000-62}).

It has been demonstrated that in low dimensions the Falicov--Kimball model can effectively be
analyzed with the help of the classical
Monte Carlo (MC) method.
\cite{MiC2005-1,MiC2005-3}  The method is general
and it is possible to apply it for different lattice geometries and for
arbitrary fillings. The most important aspect is that the model allows
for investigation of much larger systems than the original Hubbard model does.
This is of crucial importance in the presence of magnetic field that results in a formation of the cyclotron orbits. It is clear that diameters of the orbits 
should not exceed the linear system size. Consequently, in the case of the  Hubbard model one can investigate only extremely high magnetic fields when the magnetic
flux through the lattice cell is of the order of the flux quantum. This limitation 
is significantly relaxed in the case of the Falicov--Kimball model.

In the present paper we investigate a two--dimensional Falicov--Kimball 
model in a presence of perpendicular magnetic field. Since we analyze
the spinless Falicov--Kimball model, the Zeeman term is absent. The effect of the
Zeeman splitting was analyzed in, e.g., Refs. \cite{zeeman}.
For $U=0$ the model reduces itself to the Azbel--Hofstadter 
model, the solutions of which form the famous Hofstadter butterfly. 
The solutions of the $U\ne 0$ case give important guidelines how the electronic correlations 
modify the fine structure of Hofstadter butterfly.

The outline of the paper is as follows. Section 2 briefly describes a
model and a variation of the Monte Carlo method which is used to study
systems with both classical and quantum degrees of freedom. In Section
3 we present results obtained for Falicov--Kimball model in an external,
perpendicular magnetic field. Section 4 contains summary and
conclusions.

\section{Model and Computational Method}
In our study we analyze the extended Falicov--Kimball model, described by the
following Hamiltonian:
\begin{equation}
  \label{eq:ham1}
  \mathcal{H}=-\sum_{\langle i,j \rangle}t_{ij}(\vec{A}) c^\dagger_i c_j + U\sum_j f^\dagger_if_i c^\dagger_i c_i, 
\end{equation}
where $t_{ij}(\vec{A})$ is the hopping integral depending on the magnetic
field through the Peierls phase factor:
\begin{equation}
  \label{eq:t_ij}
 t_{ij}(\vec{A}) = t\exp \left( \frac{e i}{\hbar} \int_{R_i}^{R_j} \vec{A}\cdot \vec{dr}
 \right),
\end{equation}
and $\vec{A}=B(-ay,(1-a)x,0)$ is the vector potential with the parameter
$a\in [0,1]$ that allows one to distinguish between the symmetric
gauge ($a=1/2$) and the Landau gauge ($a=0$). In the numerical
calculations we have used the symmetric gauge. The same Hamiltonian
was used by Gruber {\em et al.} to analyze flux phases in the Falicov--Kimball
model.\cite{flux}

In our simulations we use a modified Metropolis algorithm.\cite{MiC2005-3} As our
system contains both itinerant fermions and localized particles, we use the
grand canonical partition function in the following form:
\begin{equation}
  \label{eq:partFn}
  \mathcal{Z}=\sum_{\mathcal{C}}\Tr_e e^{-\beta[\mathcal{H(C)}-\mu \hat{N}]},
\end{equation}
where $\mathcal{C}$ describes configuration of the localized states,
$\beta$ is the inverse temperature and $\hat{N}$ is the operator of
total number of itinerant fermions.  For a given configuration
$\mathcal{C}$ the Hamiltonian $\mathcal{H(C)}$ can be diagonalized
numerically and summation over fermionic degrees of freedom gives:
\begin{equation}
 \mathcal{Z}=\sum_{\mathcal{C}} \prod_n \left\{1+e^{-\beta[E_n(\mathcal{C})-\mu]}\right\},
\end{equation}
where $E_n(\mathcal{C})$ is $n$--th eigenenergy of $\mathcal{H(C)}$.
Introducing the free energy of the mobile particles:
\begin{equation}
\label{eq:free_en}
  \fec=-\frac{1}{\beta}\sum_n \ln \left\{ 1 + e^{-\beta[E_n(\mathcal{C})-\mu]}\right\}
\end{equation}
the partition function can be written in a form analogous to that used
for the Ising model:
\begin{equation}
\label{partfn2}
  \mathcal{Z}=\sum_{\mathcal{C}} e^{-\beta \fec},
\end{equation}
where the difference from the Metropolis algorithm is that we use the
electronic free energy instead of the internal energy.  MC simulations
allow us to estimate the partition function and thermodynamic
functions such as the specific heat. Position of the peak in the
specific heat and in the charge density wave (CDW) susceptibility allows one 
to determine the temperature $T_c$ of the transition between the ordered and 
disordered phases.\cite{MiC2005-1}

In the present paper we restrict our considerations to a special case
of half-filling both for itinerant and localized states. We have
performed simulations of on square lattices with sizes
from $10\times10$ to $30\times 30$. Here, we present results for
a $20\times20$ lattice. Although, for such a system we are unable to analyze 
the very fine details of the energy spectrum, it is possible to obtain the general
structure of the Hofstadter butterfly within a reasonable computation time. 

\section{ Results}
\begin{figure}[htb]
\includegraphics[width=.48\textwidth]{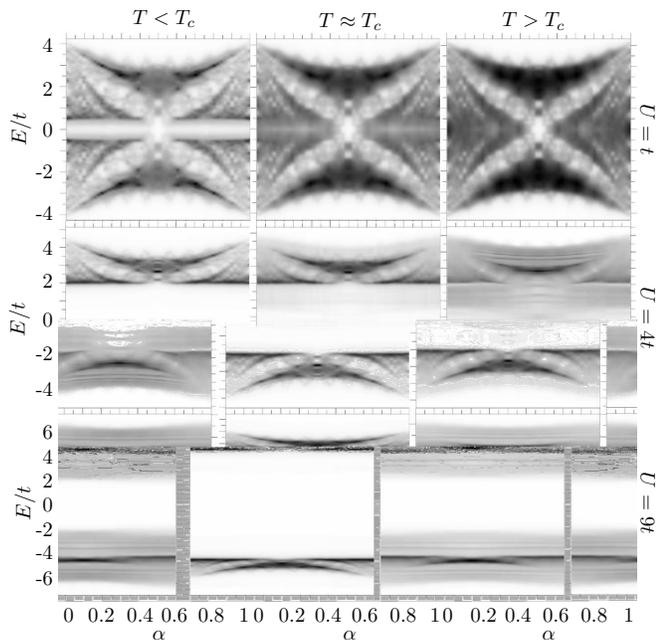}
\caption{Hofstadter butterflies for different $U/t$ and
  $k_BT/t$. Figures were obtained for a lattice of $20\times20$ sites in half
  filling for both itinerant and localized particles. Darker color
  corresponds to higher value of the density of states.}
\label{fig1}
\end{figure}

In order to the investigate thermodynamic properties of the
Falicov--Kimball model with orbital effects due to magnetic field we
performed simulations for different values of the interaction strength
$U/t$ and for the entire range of the magnetic flux penetrating the
lattice. It is convenient to use a dimensionless quantity
$\alpha=\phi/\phi_0$, where $\phi_0$ is the flux quantum and
$\alpha\in(0,1)$. It was shown by Kennedy and Lieb \cite{KennedyAndLieb}
that at low temperature the localized particles in the half--filled 
two--dimensional Falicov--Kimball model form a checkerboard pattern. At 
the same time the itinerant particles show the CDW ordering.
To reveal the most interesting properties of the
model we have performed simulations for temperatures below, near and
above the critical temperature $T_c$ of the transition between ordered and
disordered phase. The critical temperature was determined from the
peak in the specific heat versus temperature plot. The main results
are presented in Fig. \ref{fig1}.
\begin{figure}[htb]
\includegraphics[width=.46\textwidth]{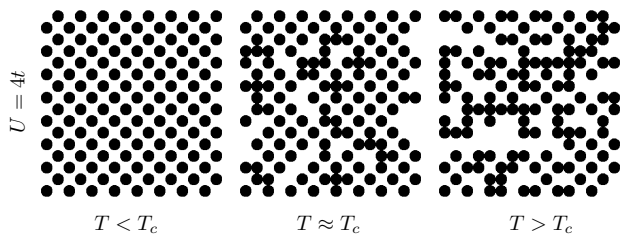}
\caption{Examples of configurations of the localized particles 
at different temperatures in the absence of magnetic field.
}
\label{fig1a}
\end{figure}
In the case of weak interactions, when
$U\le t$, the general structure of the Hofstadter butterfly is nicely
reproduced. At temperatures below $T_c$ the structure is similar to
the energy spectrum of the noninteracting fermions, but is split
near the Fermi energy by the band gap. With an increase of the temperature
the gap diminishes and vanishes completely at the $T_c$. Above $T_c$
the structure of the field dependent density of states looks similar to
the Hofstadter butterfly, but is smeared because of averaging over
disordered states. The effects of temperature on the configuration of
the localized particles is illustrated in Fig. \ref{fig1a}. The presented
configurations are ``snapshots'' of the evolution of the system during
a MC run. 
Higher interaction strengths impact the energy
structure more strongly. While for the intermediate interaction strengths
with $U\approx 4t$ at temperatures below $T_c$ the energy spectrum
still reproduces the Hofstadter butterfly split by the band gap,
increasing the temperature no longer reproduces the structure
appearing for free fermion gas. Again at the temperature of $T_c$ the
gap vanishes, but for higher temperatures the energy spectrum seems to be
weakly dependent on the magnetic field. For a strong Coulomb interaction
with $U\ge8t$ the gap persists at arbitrary temperature and its
width is not smaller than $U-8t$. An important fact is that at zero temperature 
in the absence of the interaction the magnetic field switches the system between
an insulating (or half--metallic in an infinite system) and a metallic phase,
what results from the structure of the noninteracting Hofstadter butterfly.
In the presence of the interaction this
behavior is not always reproduced. At temperatures below $T_c$ for
all analyzed values of the ratio $U/t$ the sample is always in the insulating
state, same as for $U>8t$.  On the other hand, when the temperature is
above $T_c$ for weak and intermediate interaction $U\le2t$
the magnetic field switches the system from metallic to insulating phase
(Fig. \ref{fig2}).
\begin{figure*}[htb]
\includegraphics*[width=1\textwidth]{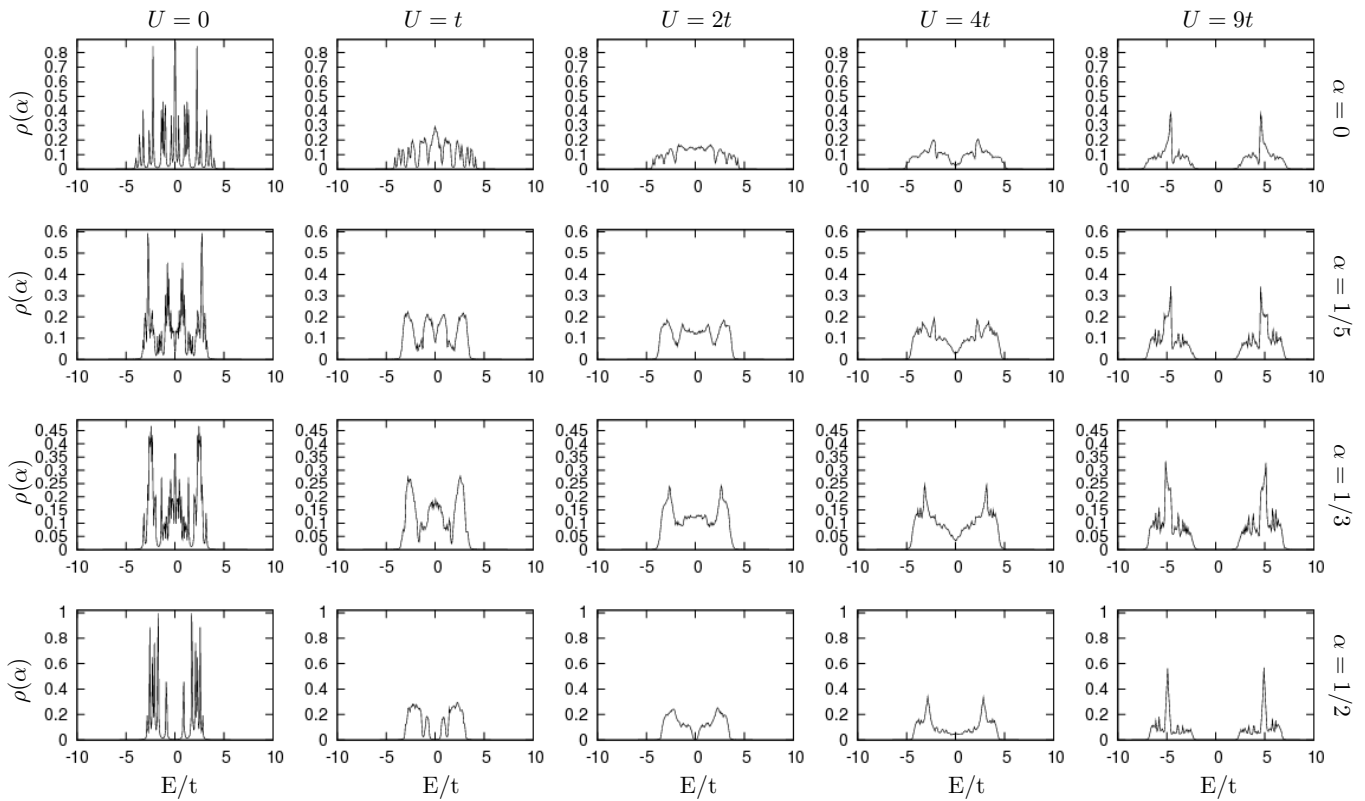}
\caption{Density of states obtained for various values of the magnetic
  flux $\alpha$ and Coulomb repulsion $U/t$. All densities of states
  were obtained from simulation at temperature $T\approx 1.5\,T_c$,
  for a lattice of $10\times10$ sites with fixed boundary conditions.}
\label{fig2}
\end{figure*}

\section{Conclusions and outlook}
We have shown that the numerical analysis of the 
Falicov--Kimball model in external magnetic field allows one to investigate 
the impact of the electron correlations on the Hofstadter butterfly. 
We investigated 
thermodynamics of the model with the Monte Carlo algorithm for
various values of Coulomb interaction, magnetic flux and
temperature. Simulations were carried out for lattices of $10\times10$
to $30\times 30$ sites for half filling for both kinds of
particles. For all probed values of $U$ at temperatures below the
$T_c$ states are insulating independently of magnetic field. For
high values of the interaction strength $U\ge8t$ samples seem to
be insulating at arbitrary temperature. An interesting result is that for
weak and intermediate interaction strength the presence and magnitude of the
energy gap depends on the magnetic field in irregular manner, as
in the noninteracting case. We show that for all investigated values of
Coulomb repulsion there is a significant smearing of the fine 
fractal structure of the energy spectrum. It occurs due to a 
broadening of quasiparticle levels. 
Nevertheless, it seems that the main branches of the Hofstadter
butterfly survive (though split by the energy gap) in the presence of 
Coulomb interactions. 

The same analysis can be repeated for the Falicov--Kimball model away from 
half filling. In this case at low temperature the localized particles 
form patterns which are incommensurate with the underlaying lattice,\cite{nne05} 
what results in a rich structure of the density of states even without 
magnetic field. The fractal structure of the Hofstadter butterfly is a result
of an interplay between two length scales: lattice constant and the Landau 
radius. In the case of the Falicov--Kimball model away from half filling
there is an additional length scale, namely the period of the pattern
formed by the localized particles. Therefore, one may expect the structure of 
the density of states versus the flux to be even richer than the 
Hofstadter butterfly.

Another extension of the present approach would be to take into account 
the spins of the itinerant and/or localized particles. It would lead to the
Zeeman splitting of the energy levels, what in turn would result in 
drastic and nontrivial changes of the density of states at the Fermi 
level.\cite{reentrant} Therefore, in such a case the simple description of the
conditions for the metal--insulator transition presented in the previous 
section would not be valid any more.

\acknowledgements
M.M.M. acknowledges a support by the Polish Ministry of Science and Higher Education
under Grant No. NN~202~128736.

%
%
\providecommand{\WileyBibTextsc}{}
\let\textsc\WileyBibTextsc
\providecommand{\othercit}{}
\providecommand{\jr}[1]{#1}
\providecommand{\etal}{~et~al.}

\end{document}